\documentclass[a4paper,twoside]{article}

\usepackage{epsfig}
\usepackage{subcaption}
\usepackage{calc}
\usepackage{amssymb}
\usepackage{amstext}
\usepackage{amsmath}
\usepackage{amsthm}
\usepackage{multicol}
\usepackage{url}
\usepackage{pslatex}
\usepackage{apalike}
\usepackage{SCITEPRESS}     

\begin{document}

\title{A Comparative Evaluation of Visual and Natural Language Question Answering Over Linked Data}

 \author{\authorname{Gerhard Wohlgenannt\sup{1}\orcidAuthor{0000-0001-7196-0699}, Dmitry Mouromtsev\sup{1}\orcidAuthor{0000-0002-0644-9242}, Dmitry Pavlov\sup{2}, Yury Emelyanov\sup{2} and Alexey Morozov\sup{2}}
 \affiliation{\sup{1}Faculty of Software Engineering and Computer Systems, ITMO University, St. Petersburg, Russia}
 \affiliation{\sup{2}Vismart Ltd., St. Petersburg, Russia}
 \email{gwohlg@corp.ifmo.ru, mouromtsev@mail.ifmo.ru,\\ \{dmitry.pavlov, yury.emelyanov, alexey.morozov\}@vismart.com}
 }


\abstract{
With the growing number and size of Linked Data datasets, it is crucial to make the data accessible and useful for users without knowledge of formal query languages. Two approaches towards this goal are knowledge graph visualization and natural language interfaces. Here, we investigate specifically question answering (QA) over Linked Data by comparing a diagrammatic visual approach with existing natural language-based systems. Given a QA benchmark (QALD7), we evaluate a visual method which is based on iteratively creating diagrams until the answer is found, against four QA systems that have natural language queries as input. Besides other benefits, the visual approach provides higher performance, but also requires more manual input. The results indicate that the methods can be used complementary, and that such a combination has a large positive impact on QA performance,
and also facilitates additional features such as data exploration. 
}

\onecolumn \maketitle \normalsize \setcounter{footnote}{0} \vfill

\section{\uppercase{Introduction}}
\label{sec:intro}

\noindent The Semantic Web provides a large number of structured datasets in form of Linked Data.
One central obstacle is to make this data available and consumable to lay users without knowledge of formal query languages such as SPARQL.
In order to satisfy specific information needs of users, a typical approach are natural language interfaces to
allow question answering over the Linked Data (QALD) by translating user queries into SPARQL~\cite{Diefenbach2018,Lopez2013}.

As an alternative method,~\cite{wohlgenannt2018eswc} propose a visual method of QA using an iterative diagrammatic approach.
The diagrammatic approach relies on the visual means only, it requires more user interaction than natural language QA,
but also provides additional benefits like intuitive insights into dataset characteristics, or a broader understanding
of the answer and the potential to further explore the answer context, and finally allows for knowledge sharing by storing and sharing resulting diagrams.

\begin{figure*}[htb]
\centering
{\centering \resizebox*{0.85\textwidth}{!}{
\includegraphics{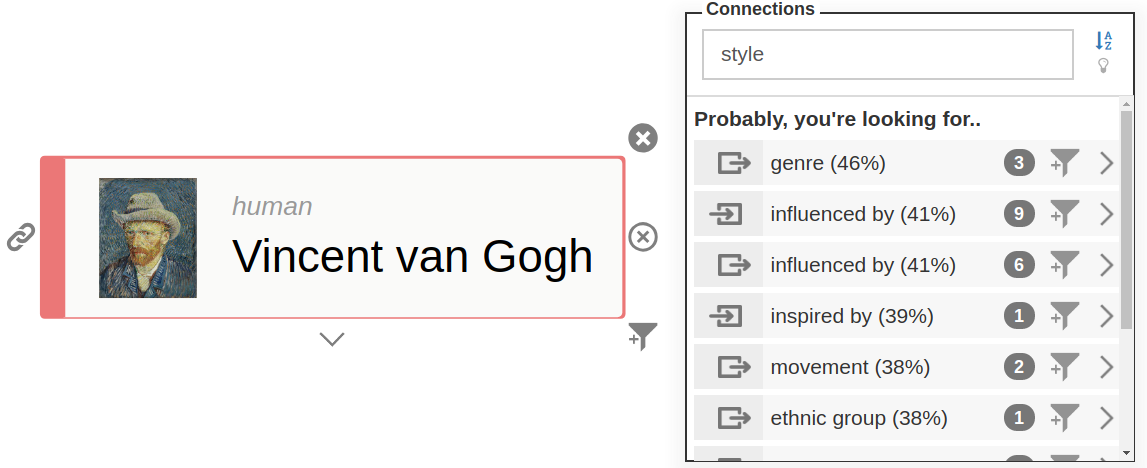}
}} 
\caption{\label{fig1} After placing the Wikidata entity \emph{Van Gogh} onto the canvas, searching properties related to his ``style'' with Ontodia DQA tool.}
\end{figure*}

In contrast to~\cite{wohlgenannt2018eswc}, who present the basic method and tool for
diagrammatic question answering (DQA), here we evaluate DQA in comparison to natural language QALD systems.
Both approaches have different characteristics, therefore we see them as complementary rather than in competition.

The basic research goals are: i) Given a dataset extracted from the QALD7 benchmark\footnote{\url{https://project-hobbit.eu/challenges/qald2017}},
we evaluate DQA versus state-of-the-art QALD systems. ii) More specifically, we investigate if and to what extent DQA
can be complementary to QALD systems, especially in cases where those systems do not find a correct answer. iii) Finally, we 
want to present the basic outline for the integration of the two methods.

In a nutshell, users that applied DQA found the correct answer with an F1-score of 79.5\%,
compared to a maximum of 59.2\% for the best performing QALD system. Furthermore, for the subset of questions where the QALD system could not provide a correct answer,
users found the answer with 70\% F1-score with DQA. 
We further analyze the characteristics of questions where the QALD or DQA, respectively, approach is better suited.

The results indicate, that aside from the other benefits of DQA, it can be a valuable component for integration
into larger QALD systems, in cases where those systems cannot find an answer, or when the user wants to explore the
answer context in detail by visualizing the relevant nodes and relations. Moreover, users can verify answers given by a QALD system using DQA in case of doubt. 

This publication is organized as follows: After the presentation of related work in Section~\ref{sec:related}, and a brief system description of the DQA tool in Section~\ref{sec:system}, 
the main focus of the paper is on evaluation setup and results of the comparison of DQA and QALD, including a discussion, in Section~\ref{sec:evaluation}.
The paper concludes with Section~\ref{sec:concl}.

\section{\uppercase{Related Work}}
\label{sec:related}


    As introduced in~\cite{wohlgenannt2018eswc} we understand diagrammatic question answering (DQA) 
    as the process of QA relying solely on visual exploration using diagrams as a representation 
    of the underlying knowledge source. 
    The process includes (i) a model for diagrammatic representation of semantic data which supports data interaction using 
    embedded queries, (ii) a simple method for step-by-step construction of diagrams with respect to cognitive boundaries and a layout that boosts understandability of diagrams,
    (iii) a library for visual data exploration and sharing based on its internal data model,
    and (iv) an evaluation of DQA as knowledge understanding and knowledge sharing tool.
    \cite{eppler2007visual} propose a framework of five perspectives of knowledge visualization,
    which can be used to describe certain aspects of the DQA use cases, such as its goal to provide
    an iterative exploration method, which is accessible to any user, the possibility of knowledge sharing
    (via saved diagrams), or the general purpose of knowledge understanding and abstraction from technical details.

    Many tools exist for visual consumption and interaction with RDF knowledge bases, however, they are
    not designed specifically towards the question answering use case.
    \cite{dudavs2018ontology} give an overview of ontology and Linked Data visualization tools,
    and categorize them based on the used visualization methods, interaction techniques and supported ontology constructs.
%


\begin{figure*}[htb]
\centering
{\centering \resizebox*{0.90\textwidth}{!}{
\includegraphics{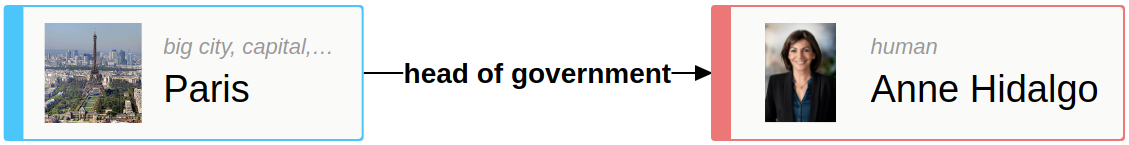}
}} \caption{\label{fig2} Answering the question: \emph{Who is the mayor of Paris?}}
\end{figure*}

        Regarding language-based QA over Linked Data, 
        \cite{kaufmann2007} discuss and study the usefulness of natural language 
        interfaces to ontology-based knowledge bases in a general way. 
        They focus on usability of such systems for the end user, and
        conclude that users prefer full sentences for query formulation and
        that natural language interfaces are indeed useful.

        \cite{Diefenbach2018} describe the challenges of QA over knowledge bases using natural languages,
        and elaborate the various techniques used by existing QALD systems to overcome those challenges. 
        In the present work, we compare DQA with four of those systems using a subset of questions of the QALD7 benchmark. Those systems are:
        gAnswer~\cite{Zou2014} is an approach for RDF QA that has a ``graph-driven'' perspective. 
        In contrast to traditional approaches, which first try to understand the question, and then evaluate the query,
        in gAnswer the intention of the query is modeled in a structured way, which leads to a subgraph matching problem.
        Secondly, QAKiS~\cite{Cabrio2014} is QA system over structured knowledge bases such as DBpedia 
        that makes use of relational patterns which capture different ways to express a certain relation in a natural language
        in order to construct a target-language (SPARQL) query.
        Further, Platypus~\cite{pellissier2018} is a QA  system on Wikidata. It represents questions in an internal format related to dependency-based compositional semantics
        which allows for question decomposition and language independence. 
        The platform can answer complex questions in several languages by using hybrid grammatical and template-based techniques. 
        And finally, also the WDAqua~\cite{Diefenbach2018} system aims for language-independence and for being agnostic of the underlying knowledge base.
        WDAqua puts more importance on word semantics than on the syntax of the user query, and follows a processes of query expansion, SPARQL construction,
        query ranking and then making an answer decision. 

        For the evaluation of QA systems, several benchmarks have been proposed 
        such as WebQuestions~\cite{berant2013semantic} or SimpleQuestions~\cite{bordes2015large}.
        However, the most popular benchmarks in the Semantic Web field arise from the QALD evaluation campaign~\cite{Lopez2013}.
        The recent QALD7 evaluation campaign includes task 4: ``English question answering over Wikidata''\footnote{\url{https://project-hobbit.eu/challenges/qald2017/qald2017-challenge-tasks/#task4}}
        which serves as basis to compile our evaluation dataset.

\section{\uppercase{System Description}}

\label{sec:system}

\noindent The DQA functionality is part of the Ontodia\footnote{\url{http://ontodia.org}} tool.
The initial idea of Ontodia was to enable the exploration of semantic graphs for ordinary users.
Data exploration is about efficiently extracting knowledge from data 
even in situations where it is unclear what is being looked for exactly~\cite{Idreos2015}. 

The DQA tool uses an incremental approach to exploration typically starting from a very small number of nodes.
With the context menu of a particular node, relations and related nodes can be added until the diagram fulfills the information need of the user.
Figure~\ref{fig1} gives an example of a start node, where a user wants to learn more about the painting style of \emph{Van Gogh}.

To illustrate the process, we give a brief example here. More details about the DQA tool, the motivation for DQA and diagram-based visualizations
are found in previous work~\cite{wohlgenannt2018eswc,wohlgenannt2017nliwod}.

As for the example, when attempting to answer a question such as ``Who is the mayor of Paris?'' the first step for a DQA user 
is finding a suitable starting point, in our case the entity \emph{Paris}. The user enters ``Paris'' into the search box,
and can then investigate the entity on the tool canvas. The information about the entity stems from the underlying dataset, for example Wikidata\footnote{\url{https://www.wikidata.org}}.
The user can -- in an incremental process -- search in the properties of the given entity (or entities) and add relevant entities onto the canvas. 
In the given example, the property ``head of government'' connects the mayor to the city of Paris, \emph{Anne Hidalgo}. 
The final diagram which answers the given question is presented in Figure~\ref{fig2}.


\section{\uppercase{Evaluation}}
\label{sec:evaluation}

\noindent Here we present the evaluation of DQA in comparison to four QALD systems.

\subsection{Evaluation Setup}

     \begin{table*}[htb]
     \vspace{-0.2cm}
     \caption{Overall performance of DQA and the four QALD tools -- measured with precision, recall and F1 score.} 
     \label{tab1}
     \centering
     \begin{center}
     \begin{tabular}{|c|c|cccc|}
     \hline
     ~~        &  DQA ~~& ~~WDAqua~~ & ~~askplatyp.us~~ &~~ QAKiS~~ &~~ gAnswer  \\ \hline
     Precision & 80.1\% & 53.7\%     & 8.57\%           & 29.6\%    & 57.5\%  \\ 
     Recall    & 78.5\% & 58.8\%     & 8.57\%           & 25.6\%    & 61.1\%  \\ \hline
     F1        & 79.5\% & 56.1\%     & 8.57\%           & 27.5\%    & 59.2\%  \\ \hline

     \end{tabular}
     \end{center}
     \end{table*}


\noindent As evaluation dataset, we reuse questions from the QALD7 benchmark task 4 ``QA over Wikidata''.
Question selection from QALD7 is based on the principles of question classification in QA~\cite{moldovan2000structure}.
Firstly, it is necessary to define question types which correspond to different scenarios of data exploration in DQA, 
as well as the type of expected answers and the question focus.
The question focus refers to the main information in the question which help a user find the answer.
We follow the model of~\cite{riloff2000rule} who categorize questions by their question word into
WHO, WHICH, WHAT, NAME, and HOW questions.
Given the question and answer type categories, we created four questionnaires with nine questions each\footnote{\url{https://github.com/ontodia-org/DQA/wiki/Questionnaires1}}
resulting in 36 questions from the QALD dataset.
The questions were picked in equal number for five basic question categories.

20 persons participated in the DQA evaluation -- 14 male and six female from eight different countries. 
The majority of respondents work within academia, however seven users were employed in industry.
131 diagrams (of 140 expected) were returned by the users.

The same 36 questions were answered using four QALD tools:
WDAqua\footnote{\url{http://qanswer-frontend.univ-st-etienne.fr}}~\cite{Diefenbach2018},
QAKiS\footnote{\url{http://qakis.org}}~\cite{Cabrio2014},
gAnswer\footnote{\url{http://ganswer.gstore-pku.com}}~\cite{Zou2014}
and Platypus\footnote{\url{https://askplatyp.us}}~\cite{pellissier2018}.

For the QALD tools, a human evaluator pasted the questions as is into the natural language Web interfaces, and submitted them to the systems.
Typically QALD tools provide a distinct answer, which may be a simple literal, or a set of entities which represent the answer,
and which can be compared to the gold standard result.
However, the \emph{WDAqua} system, sometimes, additionally to the \emph{direct answer} to the question,
provides links to documents related to the question. We always chose the answer available via \emph{direct answer}.

To assess the correctness of the answers given both by participants in the DQA experiments, and by the QALD system, 
we use the classic information retrieval metrics of precision (P), recall (R), and F1.
$P$ measures the fraction of relevant (correct) answer (items) given versus all answers (answer items) given. 
$R$ is the faction of correct answer (parts) given divided by all correct ones in the gold answer, 
and $F1$ is the harmonic mean of $P$ and $R$.
As an example, if the question is ``Where was Albert Einstein born?'' (gold answer: ``Ulm''), and the system gives 
two answers ``Ulm'' and ``Bern'', then $P=\frac{1}{2}$, $R=1$ and ${F1}=\frac{2}{3}$.


For DQA four participants answered each question, therefore we took the average $P$, $R$, and $F1$ values over the four evaluators as the result per question.
The detailed answers by the participants and available online\footnote{\url{https://github.com/ontodia-org/DQA/wiki/Experiment-I-results}}.


\subsection{Evaluation Results and Discussion}

\noindent Table~\ref{tab1} presents the overall evaluation metrics of DQA, and the four QALD tools studied.
With the given dataset, \emph{WDAqua} (56.1\% F1) and \emph{gAnswer} (59.2\% F1) clearly outperform \emph{askplatyp.us} (8.6\% F1) and \emph{QAKiS} (27.5\% F1).
Detailed results per question including the calculation of $P$, $R$ and $F1$ scores are available online\footnote{\url{https://github.com/gwohlgen/DQA_evaluations/blob/master/nlp_eval.xlsx}}.
DQA led to 79.5\% F1 (80.1\% precision and 78.5\% recall).

\begin{figure*}[htb]
\centering
{\centering \resizebox*{0.90\textwidth}{!}{
\includegraphics{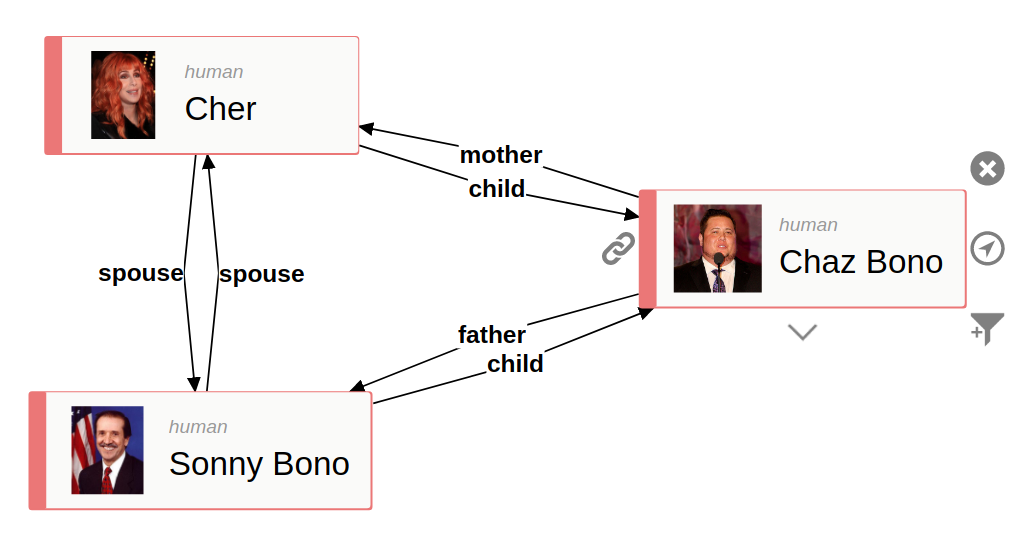}
}} \caption{\label{fig:chaz} Answering the question: \emph{Who is the son of Sonny and Cher?} with DQA.}
\end{figure*}

In further evaluations, we compare DQA results to WDAqua in order to study the differences and potential complementary aspects of the approaches.
We selected WDAqua as representative of QALD tools, as it provides state-of-the-art results, and is well grounded in the Semantic Web community.
\footnote{Furthermore, at the time of paper writing the gAnswer online demo was not available any more, support for this tools seems limited.}

Comparing DQA and WDAqua, the first interesting question is: To what extend is DQA helpful on questions that could not be answered by the QALD system?
For WDAqua the overall F1 score on our test dataset is $56.1\%$.
For the subset of questions where WDAqua had no, or only a partial, answer, DQA users found the correct answer in $69.6\%$ of cases.
On the other hand, the subset of questions that DQA users (partially) failed to answer, were answered correctly by WDAqua with an F1 of $27.3\%$.
If DQA is used as a backup method for questions not correctly answered with WDAqua, then overall F1 can be raised to $85.0\%$.
The increase from $56.1\%$ to $85.0\%$ demonstrates the potential of DQA as complementary component in QALD systems.

As expected, questions that are difficult to answer with one approach are also harder for the other approach -- as some questions in the dataset
or just more complex to process and understand than others.
However, almost 70\% of questions not answered by WDAqua could still be answered by DQA.
As examples of cases which are easier to answer for one approach than the other, a question that DQA users could answer, but where WDAqua failed is:
``What is the name of the school where Obama's wife studied?''. This complex question formulation is hard to interpret correctly for a machine.
In contrast to DQA, QALD systems also struggled with ``Who is the son of Sonny and Cher?''. This question needs a lot of real-world knowledge to map the
names \emph{Sonny} and \emph{Cher} to their corresponding entities. The QALD system needs to select the correct \emph{Cher} entity from multiple options
in Wikidata, and also to understand that ``Sonny'' refers to the entity \emph{Sonny Bono}.
The resulting answer diagram is given in Figure~\ref{fig:chaz}.
More simple questions, like ``Who is the mayor of Paris?'' were correctly answered by WDAqua, but not by all DQA users. DQA participants in this case
struggled to make the leap from the noun ``mayor'' to the \emph{head-of-government} property in Wikidata.

Regarding the limits of DQA, this method has difficulties when the answer can be obtained only with joins of queries,
or when it is hard to find the initial starting entities related to question focus.
For example, a question like ``Show me the list of African birds that are extinct.'' typically requires an intersection of two (large) sets of candidates entities,~ie. all \emph{African birds} and \emph{extinct birds}.
Such a task can easily be represented in a SPARQL query, but is hard to address with diagrams, because it would require placing, and interacting with, a huge amount of nodes on the exploration canvas.

Overall, the experiments indicate, that additionally to the use cases where QALD and DQA are useful on their own,
there is a lot of potential in combining the two approaches, especially by providing a user the opportunity to
explore the dataset with DQA if QALD did not find a correct answer, or when a user wants to confirm the QALD answer by checking in the underlying knowledge base.
Furthermore, visually exploring the dataset provides added benefits, like understanding the dataset characteristics,
sharing of resulting diagrams (if supported by the tool), and finding more information related to the original information need.

For the integration of QALD and DQA, we envision two scenarios.
The first scenario addresses plain question answering, and here DQA
can be added to a QALD system for cases where a user is not satisfied with a given answer. The QALD Web interface can
for example have a \emph{Explore visually with diagrams} button, which brings the user to a canvas on which the
entities detected by the QALD system within the question and results (if any) are displayed on the canvas as starting nodes.
The user will then explore the knowledge graph and find the answers in the same way as the participants in our experiments.
The first scenario can lead to a large improvement in answer F1 (see above).

The second scenario of integration of QALD and DQA focuses on the exploration aspect. Even if the QALD system provides the correct answer,
a user might be interested to explore the knowledge graph to validate the result and to discover more interesting information about the target entities.
From an implementation and UI point of view, the same \emph{Explore visually with diagrams} button and pre-population of the canvas can be used.
Both scenarios also provide the additional benefits of potentially saving and sharing the created diagrams, which elaborate the relation between
question and answer.




\section{\uppercase{Conclusions}}
\label{sec:concl}

\noindent In this work, we compare two approaches to answer questions over Linked Data datasets: a visual diagrammatic approach
(DQA) which involves iterative exploration of the graph, and a natural language-based (QALD).
The evaluations show, that DQA can be a helpful addition to pure QALD systems, both regarding evaluation metrics (precision, recall, and F1), and also for dataset understanding and further exploration.
The contributions include: i) a comparative evaluation of four QALD tools and DQA with a dataset extracted from the QALD7 benchmark,
ii) an investigation into the differences and potential complementary aspects of the two approaches, and iii) the proposition of
integration scenarios for QALD and DQA.

In future work we plan to study the integration of DQA and QALD, especially the aspect of automatically creating an initial diagram from a user query,
in order to leverage the discussed potentials. We envision an integrated tool, that uses QALD as basic method to find an answer to a question
quickly, but also allows to explore the knowledge graph visually to raise answer quality and support exploration with all its discussed benefits.

\section*{\uppercase{Acknowledgements}}
\noindent This work was supported by the Government of the Russian Federation (Grant 074-U01) through the ITMO Fellowship and Professorship Program.

\bibliographystyle{apalike}
{\small
\bibliography{here}}

\end{document}